# Macroblock Classification Method for Video Applications Involving Motions

Weiyao Lin, Ming-Ting Sun, Hongxiang Li, Zhenzhong Chen, Wei Li, and Bing Zhou

*Abstract*—**In this paper, a macroblock classification method is proposed for various video processing applications involving motions. Based on the analysis of the Motion Vector field in the compressed video, we propose to classify Macroblocks of each video frame into different classes and use this class information to describe the frame content. We demonstrate that this low-computation-complexity method can efficiently catch the characteristics of the frame. Based on the proposed macroblock classification, we further propose algorithms for different video processing applications, including shot change detection, motion discontinuity detection, and outlier rejection for global motion estimation. Experimental results demonstrate that the methods based on the proposed approach can work effectively on these applications.**

## I. INTRODUCTION AND RELATED WORK

Video processing techniques such as video compression and video content analysis have been widely used in various applications [1-17]. In many of these techniques and applications, motion-based features play an important role since they are closely related to the 'dynamic' nature of videos [1-17].

There have been many researches which use the motion-based or motion-related features for video processing. Efros et al. [4] and Chaudhry et al. [5] use optical flow to detect human and recognize their activities in video. Lin et al. [6] and Chen et al. [7] analyze the video contents by first tracking and extracting the motions of objects and humans. However, in many applications, video processing steps are often integrated with the video compression module, for example, analyzing video contents for facilitating rate control [8], detecting gradual shot changes for applying weighted motion prediction [9] for improving video compression efficiency, segmenting irregular motion regions for improving global motion estimation efficiency, and labeling shot change or motion discontinuity places during video compression for further editing. Most of these applications require the video processing algorithms to have low complexity such that few computation overheads are introduced to the computation-intensive video compression module. From this point of view, many of the above mentioned algorithms have high computation complexity and are not suitable for these applications.

Furthermore, many works also extract motion features from the Motion Vector (MV) information which is already available in many compressed-domain videos. Akutsu et al. [10] and Shu et al. [11] detect the shot changes based on the information of MV motion smoothness. However, their methods have limitations in differentiating shot changes and motion discontinuities. Porikli et al. [12] and Yoon et al. [13] utilize the compressed-domain MV field for object segmentation or event detection. Su et al. [14] utilize the MV field information to speed up the global motion estimation. Although these methods can create satisfying results, their complexities are still high when integrated with the computation-intensive video compression module. Furthermore, although some other MV-feature-based methods are proposed which try to improve the video processing performance with reduced complexity [15-17], most of their motion features only focus on one specific application and are often unsuitable when applied on other applications.

In this paper, a new Macroblock (MB) classification method is proposed which can be used for various video processing applications. According to the analysis of the MV field, we first classify the Macroblocks of each frame into different classes and use this class information to describe the frame content. Based on the proposed approach, we further propose algorithms for various video processing applications including shot change detection, motion discontinuity detection, and outlier rejection for global motion estimation. Experimental results demonstrate that algorithms based on the proposed approach can work efficiently and perform better than many existing methods. Since the proposed MB class information is extracted from the information readily available in the Motion Estimation (ME) process [2, 18] or from the compressed bit-stream, its computation overhead is low. It can easily be implemented into most existing video coding systems without extra cost.

The rest of the paper is organized as follows: Section II describes our proposed MB classification method. Based on the proposed approach, Section III proposes three algorithms for shot change detection, motion discontinuity detection, and outlier rejection for global motion estimation applications,

W. Lin is with the Institute of Image Communication and Information Processing, Department of Electronic Engineering, Shanghai Jiao Tong University, Shanghai 200240, China (email: wylin@sjtu.edu.cn).

M.-T. Sun is with the Department of Electrical Engineering, University of Washington, Seattle, WA 98195, USA (e-mail: mts@u.washington.edu).

H. Li is with the Department of Electrical and Computer Engineering, University of Louisville, Louisville, KY 40292, USA (email: hongxiang.li@gmail.com).

Z. Chen is with the School of Electrical and Electronic Engineering, Nanyang Technological University, Singapore 639798 (email: zzchen@ntu.edu.sg).

W. Li is with the School of Computer Science and Engineering, Beihang University, Beijing 100083,China (email: lwei@buaa.edu.cn).

B. Zhou is with the School of Information Engineering, Zhengzhou University, Zhengzhou 450001, China (email: iebzhou@zzu.edu.cn).

respectively. The experimental results are given in Section IV. Section V discusses some possible extensions, and Section VI summarizes the paper.

## II. THE MB CLASSIFICATION METHOD

In most practical applications, videos are processed and stored in the compressed domain where ME is performed during the compression process to remove the temporal redundancy. Since ME is a process to match similar areas between frames, much information related to frame content correlation and object motion are already available from the ME process. The compressed video provides the MV information which can be directly extracted from the bitstream. Therefore, in this section, we propose to use MV information to classify MBs.

Without loss of generality, the MB classification method can be described in Eqn. (1).

$$Class_{cur\_MB} = \begin{cases} 1 & \text{if } init\_COST < Th_1 \\ 2 & \text{if } init\_COST \geq Th_1 \text{ and } |PMV_{cur\_MB} - MV_{pre\_final}| > Th_2 \\ 3 & \text{if } init\_COST \geq Th_1 \text{ and } |PMV_{cur\_MB} - MV_{pre\_final}| \leq Th_2 \end{cases} \quad (1)$$

where $cur\_MB$ is the current MB, $init\_COST$ is the initial matching cost value calculated based on the motion information of spatial or temporal neighboring MBs, $Th_1$ is a threshold, $PMV_{cur\_MB}$ is the Predictive Motion Vector of the current MB [18], $MV_{pre\_final}$ is the final Motion Vector (MV) of the co-located MB in the previous frame, and $Th_2$ is another threshold checking the closeness between $PMV_{cur\_MB}$ and $MV_{pre\_final}$. Using Eqn. (1), MBs with small $init\_COST$ values will be classified as Class 1. MBs will be classified as Class 3 if their PMVs are close to the final MVs of their collocated MBs in the previous frame. Otherwise, MBs will be classified into Class 2. The motivation of using Eqn. (1) is that the variables involved are all readily available from most of the ME processes.

The motivations of classifying MBs according to Eqn. (1) can be summarized as follows:

(1) According to Eqn. (1), MBs in Class 1 have two features: (a) their MVs can be predicted accurately (i.e., $init\_COST$ is calculated based on the motion information of spatial or temporal neighboring MBs). This means that the motion patterns of these MBs are regular (i.e., can be predicted) and smooth (i.e., coherent with the previous-frame motions). (b) They have small matching cost values. This means that these MBs can find good matches from the previous frames. Therefore, the Class 1 information can be viewed as an indicator of the content correlation between frames.

(2) According to Eqn. (1), Class 2 includes MBs whose motion cannot be accurately predicted by their neighboring information (PMV) and their previous motion information ($MV_{pre\_final}$). This means that the motion patterns of these MBs are irregular and unsmooth from those of the previous frames. Therefore, the Class 2 information can be viewed as an indicator of the motion unsmoothness between frames.

(3) According to Eqn. (1), Class 3 includes MBs whose MVs are close to the PMVs and whose matching cost values are large. Therefore, Class 3 MBs will include areas with complex textures but similar motion patterns to the previous frames.

Fig. 1 shows two example classification results for two sequences using Eqn. (1). The experimental setting is the same as that described in Section IV. Fig. 1 (a) and (e) are the original frames. Blocks labeled grey in (b) and (f) are MBs belonging to Class 1. Blocks labeled black in (c) and (g) and blocks labeled white in (d) and (h) are MBs belonging to Class 2 and Class 3, respectively.

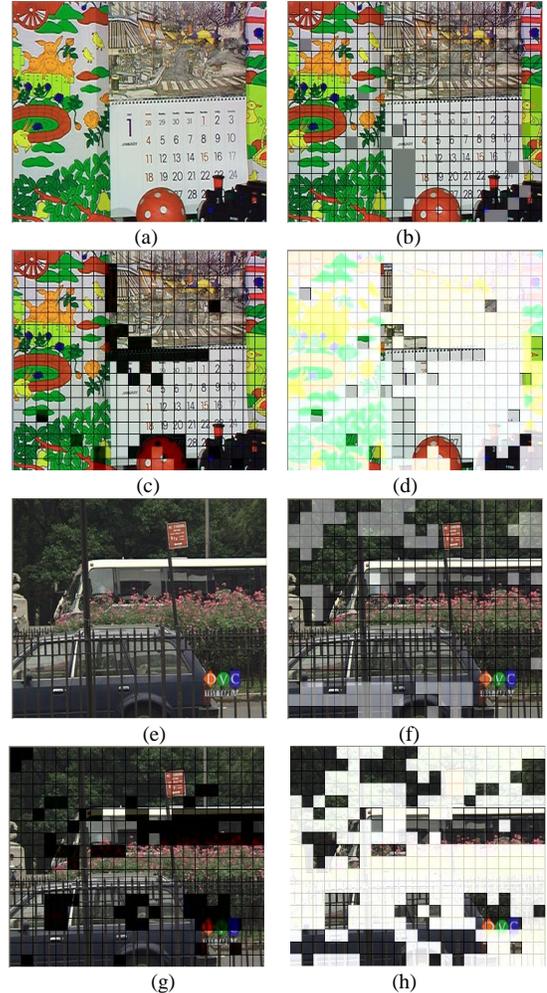

Fig. 1. The original frames (a, e) and the distributions of Class 1 (b, f), Class 2 (c, g), and Class 3 (d, h) MBs for *Mobile_Cif* and *Bus_Cif* using Eqn. (1).

Several observations can be drawn from Fig. 1 as follows:

From Fig. 1 (b) and (f), we can see that most Class 1 MBs include backgrounds or flat areas that can find good matches in the previous frames. From Fig. 1 (c) and (g), we can see that our classification method can effectively detect *irregular areas* and classify them into Class 2 (for example, the *edge between the calendar and the background* as well as *the bottom circling ball* in (c), and *the running bus* as well as *the down-right logo* in (g)). From Fig. 1 (d) and (h), we can see that most complex-texture areas are classified as Class 3, such as the complex background and calendar in (d) as well as the flower area in (h).

Since *init_COST* is only available in the ME process, Eqn. (1) is more suitable for applications where video coding and other video processing are performed at the same time, such as global motion estimation, rate control, computation control coding [8], as well as labeling shot changes in the process of compressing videos [2]. However, it should be noted that Eqn. (1) is only an implementation example of the proposed classification method. The idea of the proposed MB classification is general and it can be easily extended to other forms for different applications. For example, for some compressed-domain video processing applications (i.e. processing already-compressed videos where *init_COST* is not readily available), Eqn. (1) can be extended to Eqn. (2):

$$Class_{cur\_MB} = \begin{cases} 1 & if \ SUM_{red} < Th_1 \ and \ |PMV_{cur\_MB} - MV_{pre\_final}| \leq Th_2 \\ 2 & if \ |PMV_{cur\_MB} - MV_{pre\_final}| > Th_2 \\ 3 & if \ SUM_{red} \geq Th_1 \ and \ |PMV_{cur\_MB} - MV_{pre\_final}| \leq Th_2 \end{cases} \quad (2)$$

where $SUM_{red}$ is the absolute sum of the decoded residual of the current MB [19]. Fig. 2 shows the classification results using Eqn. (2) where *Th1* and *Th2* are set to be the same as in Fig. 1. We can see from Fig. 2 that Eqn. (2) can result in similar classification results as Eqn. (1). In the following, we will perform discussion and experiments according to Eqn. (1) in the rest of the paper.

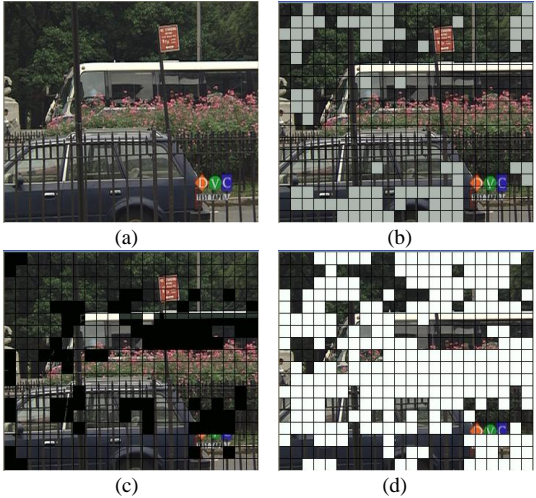

Fig. 2. The original frames (a) and the distributions of Class 1 (b), Class 2 (c), and Class 3 (d) MBs for *Mobile_Cif* using Eqn. (2).

With the proposed MB class information, we can develop various algorithms for different applications. Since our proposed method is directly defined based on the information readily available from the ME process or from the compressed video bitstream, it is with low computational complexity and is applicable to various video applications, especially for those with low-delay and low-cost requirements. In the following section, we will propose algorithms for the three example applications: shot change detection, motion discontinuity detection, and outlier rejection for global motion estimation.

## III. USING THE MB CLASS INFORMATION FOR VIDEO APPLICATIONS

### A. Shot Change Detection

In this paper, we define a 'shot' as a segment of continuous video frames captured by one camera action (i.e., a continuous operation of one camera), and a 'shot change' as the boundary of two shots [2]. Fig. 3 shows an example of an abrupt shot change.

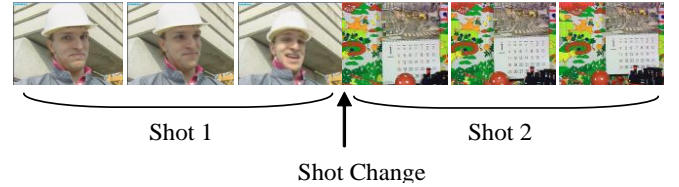

Fig. 3 An example of an abrupt shot change.

From the discussions in the previous section, we can outline the ideas of applying our approach to shot change detection as follows:

Since shot changes (including abrupt, gradual, fade-in or fade-out) [2] always happen between two uncorrelated video shots, the content correlation between frames at shot changes will be low. Therefore, we can use the information of Class 1 as the primary feature to detect shot changes. Furthermore, since the motion pattern will also change at shot changes, the information of Class 2 and Class 3 can be used as additional features for shot change detection.

Fig. 4 shows an example of the effectiveness in using our class information for shot change detection. More results will be shown in the experimental results. Fig. 4 (b)-(d) show the MB distributions of three classes at the abrupt shot change from *Bus_Cif* to *Mobile_Cif*. We can see that the information of Class 1 can effectively indicate the low content correlation between frames at the shot change (i.e., no MB is classified as Class 1 in Fig. 4-(b)). Furthermore, a large number of MBs are classified as Class 2. This indicates the rapid motion pattern change at the shot change.

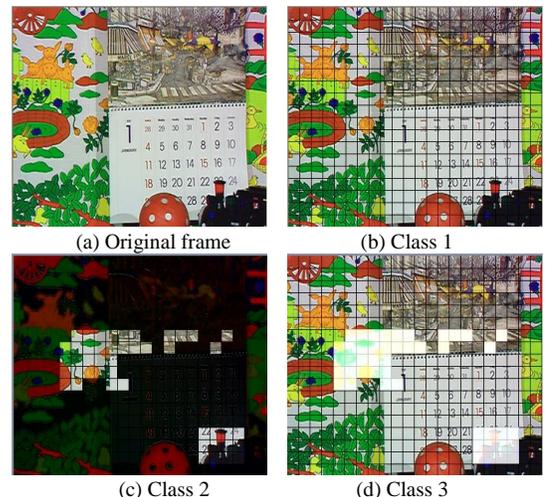

Fig. 4 The MB distributions at the abrupt shot change frame from *Bus_Cif* to *Mobile_Cif*.

Based on the above discussions, we can propose a Class-Based Shot Change detection (CB-Shot) algorithm. It is described as in Eqn. (3):

$$Fg_{shot}(t) = \begin{cases} 1 & \begin{array}{l} if \quad N_{class\_1}(t) \leq T_1 \text{ and } N_{Intra\_MB}(t) - N_{IR}(t) \geq T_4 \\ or \ if \begin{cases} N_{class\_1}(t) \leq T_2 \text{ and } N_{Intra\_MB}(t) - N_{IR}(t) \geq T_4 \text{ and} \\ |N_{class\_2}(t) - N_{class\_2}(t-1)| + |N_{class\_3}(t) - N_{class\_3}(t-1)| \geq T_3 \end{cases} \end{array} \\ 0 & else \end{cases} \quad (3)$$

where $t$ is the frame number and $Fg_{shot}(t)$ is a flag indicating whether a shot change happens at the current frame $t$ or not. $Fg_{shot}(t)$ will equal to $1$ if there is a shot change and will equal to $0$ else. $N_{Intra\_MB}(t)$ is the number of intra-coded MBs at frame $t$, $N_{IR}(t)$ is the number of intra-refresh MBs in the current frame (i.e., forced intra-coding MBs [20]). $N_{class\_1}(t)$, $N_{class\_2}(t)$ and $N_{class\_3}(t)$ are the total number of Class 1, Class 2 and Class 3 MBs in the current frame $t$, respectively. $T_1$, $T_2$, $T_3$ and $T_4$ are the thresholds for deciding the shot change. In this paper, $T_1$-$T_4$ are calculated by Eqn. (4).

$$T_1 = \frac{(N_{MB}(t) - N_{IR}(t))}{40}, T_2 = \frac{(N_{MB}(t) - N_{IR}(t))}{30}, T_3 = \frac{(N_{MB}(t) - N_{IR}(t))}{4}, T_4 = T_1 \quad (4)$$

where $N_{MB}(t)$ is the total number of MBs of all classes in the current frame.

It should be noted that in Eqn. (3) the Class 1 information is the main feature for detecting shot changes (i.e., $N_{class\_1}(t) \leq T_1$ and $N_{class\_1}(t) \leq T_2$ in Eqn. (3)). The intuitive of using the Class 1 information as the major feature is that it is a good indicator of the content correlation between frames. The Class 2 and Class 3 information is used to help detect frames at the beginning of some gradual shot changes where a large change in motion pattern has been detected but the number of Class 1 MBs has not yet decreased to a small number. The intra-coded MB information can help discard the possible false alarm shot changes due to the MB mis-classfication. From, Eqn. (3) and (4), we can also see that when intra-refresh functionality is enabled (i.e., when $N_{IR}(t) > 0$), our algorithm can be extended by simply excluding these intra-refreshed MBs and only performing shot change detection based on the remaining MBs.

Furthermore, note that Eqn. (3) is only one implementation of using our class information for shot change detection. We can easily extend Eqn. (3) by using more sophisticated methods such as cross-validation [6] to decide the threshold values in an automatic way. Besides, other machine learning models can also be used to decide the shot detection rules and to take the place of the manually-set rules in Eqn. (3). For example, we can train a support vector machine (SVM) or a Hidden Markov Model (HMM) based on our class information for detecting shot changes [21, 22]. By this way, we can avoid the tediousness of manually tuning multiple thresholds simultaneously. This point will be further discussed in the experimental results.

*B. Motion Discontinuity Detection*

We define motion discontinuity as the boundary between two Smooth Camera Motions (SCMs), where SCM is a segment of continuous video frames captured by one single motion of the camera (such as zooming, panning, or tilting) [2, 11]. For example, in Fig. 5, the first several frames are captured when the camera has no or little motion. Therefore, they form the first SCM (SCM1). The second several frames form another SCM (SCM2) because they are captured by a single camera motion of rapid rightward. Then, a Motion Discontinuity (MD) can be defined between these two SCMs. It should be noted that the major difference between shots and SCMs is that a shot is normally composed of multiple SCMs.

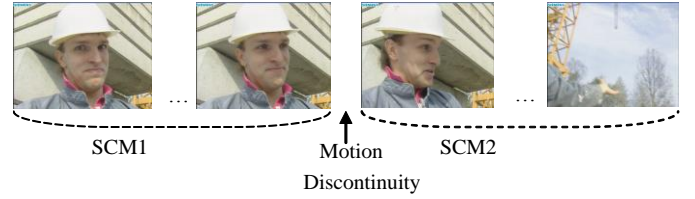

Fig. 5 An example of motion discontinuity.

Basically, motion discontinuity can be viewed as motion unsmoothness or the change of motion patterns. The detection of motion discontinuity can be very useful in video content analysis or video coding performance improvement [9, 23]. Since our class information, especially Class 2 information, can efficiently reflect the irregular motion patterns, it can be easily used for motion discontinuity detection.

The ideas of applying our MB class information into motion discontinuity detection can be outlined as follows:

Since MD happens between different camera motions, the motion smoothness will be disrupted at the places of MDs. Therefore, we can use the Class 2 information as the primary feature to detect MDs. Furthermore, since frames at MDs belong to the same camera action (i.e., the same shot), their content correlation will not decrease. Therefore, the information of Class 1 can be used to differentiate shot changes from MDs.

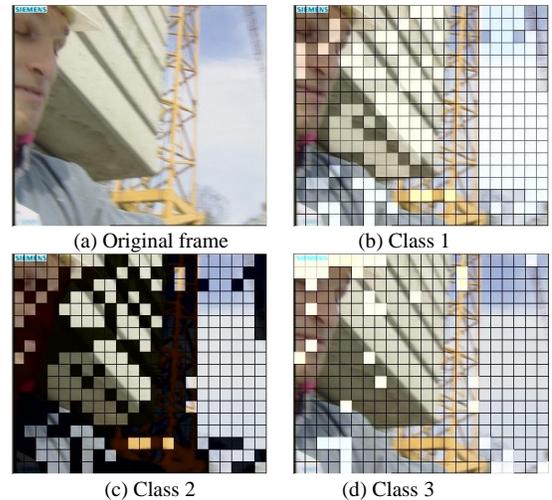

(a) Original frame     (b) Class 1
(c) Class 2     (d) Class 3

Fig. 6 The MB distributions at a Motion Discontinuity frame in *Foreman_Cif*.

Fig. 6 shows an example of the effectiveness in using our class information in MD detection. Fig. 6 (b)-(d) show the MB distributions of a Motion Discontinuity frame in *Foreman_Cif* when the camera starts to move rightward rapidly. The large number of Class 2 MBs indicates the motion unsmoothness due to the MD. Furthermore, the big number of Class 1 MBs indicates the high content correlation between frames, which implies that it is not a shot change.

Therefore, we can propose a Class-Based Motion Discontinuity Detection (CB-MD) algorithm. It is described as in Eqn. (5):

$$Fg_{MD}(t) = \begin{cases} 1 & if\ N_{class\_1}(t) \geq th_1^{MD}\ and\ \sum_{i=0}^{k} I(N_{class\_2}(t-i) \geq th_3^{MD}) = k+1 \\ 0 & else \end{cases} \quad (5)$$

where $I(f)$ is an indicator. $I$ will equal to $1$ if $f$ is true, and $0$ if $f$ is false. Eqn. (5) means that an *MD* will be detected only if the number of Class 2 MBs is larger than a threshold for $k+1$ consecutive frames. This is based on the assumption that an obvious camera motion change will affect several frames rather than one. By including the information of several frames, the false alarm rate can be reduced. Furthermore, similar to shot change detection, the decision rules in Eqn. (5) can also be extended to avoid the manual setting of thresholds.

### C. Outlier Rejection for Global Motion Estimation

Global motion estimation is another useful application of our class information. Since a video frame may often contain various objects with different motion patterns and directions, motion segmentation is needed to filter out these irregular motion regions before estimating the global motion parameters of the background. Since our class information can efficiently describe the motion patterns of different MBs, it is very useful in filtering out the irregular motion areas (outliers). For example, we can simply filter out Class-2 or Class-2+Class-3 MBs and perform global motion estimation based on the remaining MBs.

Based on the MB class information, the proposed global motion estimation algorithm can be described as follows:

**Step 1:** Use our class information to get a segmentation of the irregular motion MBs, as shown in Eqn. (6):

$$Seg_{IrregularMotion} = \begin{cases} Class2\ and\ Class3\ MBs & if\ (N_{class\_2}(t) + N_{class\_3}(t)) < Th_F \\ Class2\ MBs & else \end{cases} \quad (6)$$

where $N_{class\_2}(t)$ and $N_{class\_3}(t)$ are the number of Class 2 and Class 3 MBs in $t$, and $Th_F$ is a threshold.

**Step 2:** Estimate the global motion parameters based on the remaining background MVs. In this paper, we use the 6-parameter model as the global motion model, as described in Eqn. (7).

$$\begin{pmatrix} x' \\ y' \\ 1 \end{pmatrix} = GMV^{6p}(x,y;a,b,c,d,e,f) = S(a,b,c,d,e,f) \times \begin{pmatrix} x \\ y \\ 1 \end{pmatrix} \quad (7)$$

where $S(a,b,c,d,e,f) = \begin{pmatrix} a & b & c \\ d & e & f \\ 0 & 0 & 1 \end{pmatrix}$ is the 6-parameter model.

$(x,y)$ and $(x',y')$ represent the pixel's original and global-motion-compensated location, respectively. There are many ways to estimate $S$. In this paper, we use the Least-Square method [17] which searches parameters in $S$ that minimizes a given cost function (mean-square error), as in Eqn. (8).

$$S(a,b,c,d,e,f) = arg\ \min_{S(a,b,c,d,e,f)} \left( \sum_{x,y} \left\| V(x,y) - GMV^{6p}(x,y;a,b,c,d,e,f) \right\|^2 \right) \quad (8)$$

where $V(x,y) = (V^x\ V^y\ 1)^T$ and $(V^x, V^y)$ are the MV terminate coordinates for pixel $(x,y)$.

Fig. 7 shows some results of using our class information for irregular motion region segmentation. From Fig. 7 (a)-(b), we can see that our class information can efficiently locate the foreground object regions. However, from Fig. 7 (c), we can also see that our algorithm more focuses on detecting the "irregular motion regions" instead of the foreground object. In Fig. 7 (c), since only the person's left hand is moving while the other parts of the person keep unchanged, only those blocks corresponding to the left hand are identified as irregular motion regions.

Note that although our class information is focused on detecting irregular motion regions in this paper, it can also be extended to detect real foreground objects by combining with texture information such as DC and AC coefficients [12].

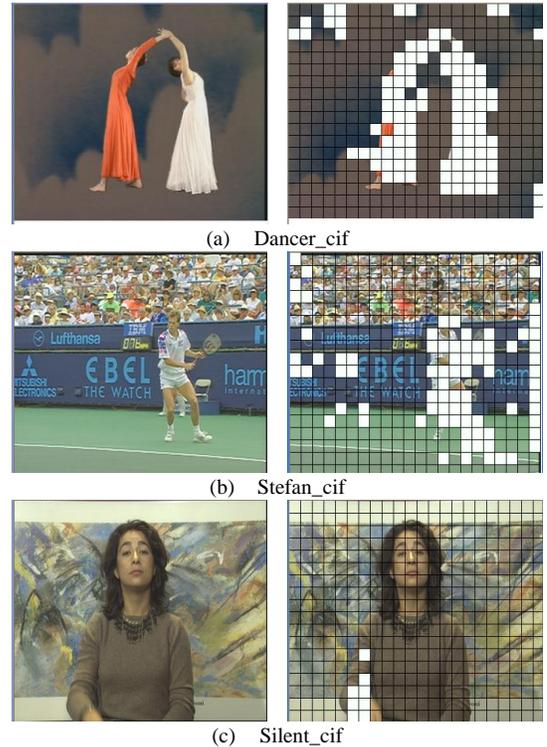

(a) Dancer_cif

(b) Stefan_cif

(c) Silent_cif

Fig. 7 Examples of using our class information for irregular motion region segmentation for global motion estimation (Left column: original frames; right column: segmented frames)

## IV. EXPERIMENTAL RESULTS

In this section, we perform experiments for the proposed methods in Section III. The algorithms are implemented on the H.264/MPEG-4 AVC reference software JM10.2 version [20]. The picture resolutions for the sequences are CIF and SIF. For each of the sequences, the picture coding structure was IPPP…. In the experiments, only the 16x16 partition was used with one reference frame coding for the P frames. The QP was set to be 28, the search range was $\pm 32$ pixels, and the frame rate was 30 frame/sec. The motion estimation is based on our proposed Class-based Fast Termination method [18]. Note that our MB classification method is general regardless of the ME algorithms used. It can easily be extended to other ME algorithms [24-25]. Furthermore, we disable the intra-refresh functionality [20] in the experiments in this paper in order to focus on our class information. However, from our experiments, the shot detection results will not differ by much when intra-refresh is enabled.

### A. Experimental Results for Shot Change Detection

We first perform experiments for shot change detection. Four shot change detection algorithms are compared.

**(1) Detect shot changes based on the number of Intra MBs** [26-27] (*Intra-based* in Table 1). A shot change will be detected if the number of Intra MBs in the current frame is larger than a threshold.

**(2) Detect shot changes based on motion smoothness** [10, 11] (*MV-Smooth-based* in Table 1). The motion smoothness can be calculated by the Square of Motion Change [11], as in Eqn. (9):

$$SMC(t) = \sum_{i \in current\_frame} \left( (MV_x^i(t) - MV_x^i(t-1))^2 + (MV_y^i(t) - MV_y^i(t-1))^2 \right) \quad (9)$$

where $SMC(t)$ is the value of the *Square of Motion Change* at frame $t$. $MV_x^i(t)$ and $MV_y^i(t)$ are the $x$ and $y$ component of the motion vector for Macroblock $i$ of frame $t$, respectively. From Eqn. (9), we can see that $SMC$ is just the 'sum of squared motion vector difference' between co-located MBs of neighboring frames. Based on Eqn. (9), a shot change can be detected if $SMC(t)$ is larger than a threshold at frame $t$.

**(3) Detect shot changes based on the combined information of Intra MB and motion smoothness** [11] (*Intra+MV-Smooth* in Table 1). In this method, the Intra-MB information is included into the *Square of Motion Change*, as in Eqn. (10).

$$SMC_{Intra\_included}(t) = \sum_{i \in current\_frame} MC(i) \quad (10)$$

where $SMC_{Intra\_included}(t)$ is the *Square of Motion Change with Intra-MB information included*. $MC(i)$ is defined as:

$$MC(i) = \begin{cases} (MV_x^i(t) - MV_x^i(t-1))^2 + (MV_y^i(t) - MV_y^i(t-1))^2 & \text{if } i \text{ is inter-coded} \\ L & \text{if } i \text{ is intra-coded} \end{cases} \quad (11)$$

where $i$ is the MB number, $L$ is a large fixed number. In the experiment of this paper, we set $L$ to be *500*. From Eqn. (10) and Eqn. (11), we can see that the *Intra+MV-Smooth* method is similar to the *MV-Smooth-based* method except that when MB $i$ is intra-coded, a large value $L$ will be used instead of the *squared motion vector difference*. It should be noted that when the number of intra MBs is low, the *Intra+MV-Smooth* method will be close to the *MV-Smooth-based* method. If the number of intra MBs is high, the *Intra+MV-Smooth* method will be close to the *Intra-based* method.

**(4) The proposed Class-Based shot change detection algorithm** which uses the Class 1 information as the major feature for detection, as in Eqn. (3) (*Proposed-All Class+Intra* in Table 1).

It should be noted that we choose Method (I)-(III) as the reference algorithms to compare with our methods because they are all computationally efficient methods (with the average operation time less than 5 ms). Thus, they are suitable for the application of shot change detection for video coding. More comparisons with other methods will also be provided in the experiment of Table 2.

Fig. 8 shows the curves of features that are used in the above algorithms. Since all the algorithms perform well in detecting abrupt shot changes, we only show the curves of a gradual shot change in Fig. 8.

Fig. 8 (a)-(e) are the feature curves of a gradual shot change sequence as in Fig. 9 (a). In this sequence, the first 5 frames are *Bus_Cif*, the last 5 frames are *Football_Cif*, and the middle 20 frames are the period of the gradual shot change. Fig. 8-(a) is the ground-truth for the shot change sequence; Fig. 8-(b) shows the curve of *the number of Intra MBs* in each frame; Fig. 8-(c) shows the curve of $SMC(t)$; Fig. 8-(d) shows the curve of $SMC_{Intra\_included}(t)$; and Fig. 8-(e) shows the curve of *the number of Class 1 MBs* in each frame. It should be noted that we reverse the $y$-axis of Fig. 8-(e) so that the curve has the same concave shape as others.

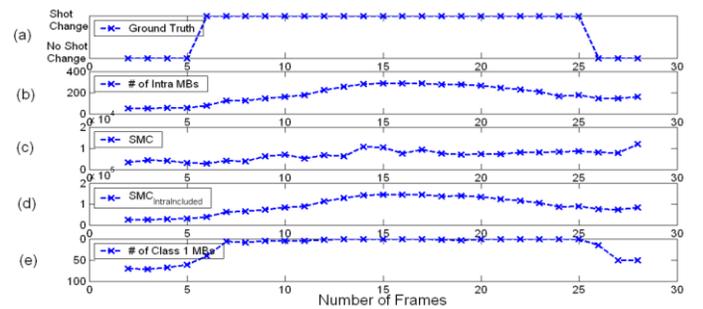

Fig. 8 Feature curves of a gradual shot change sequence.

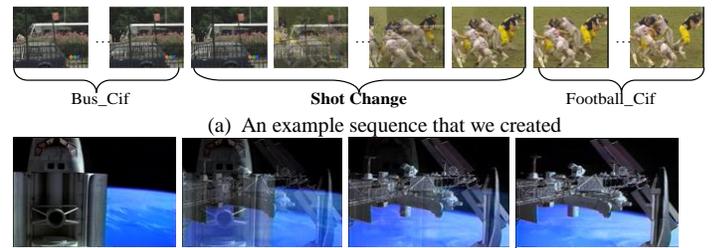

(a) An example sequence that we created

(b) The example sequence from TRECVID dataset [28]

Fig. 9 Example sequences in the extended TRACVID dataset.

Fig. 8 shows the effectiveness of using our class information for shot change detection. From Fig. 8 (e), we can see that *the number of Class 1 MBs* suddenly decreases to *0* when a shot change happens and then quickly increases to a large number right after the shot change period. Therefore, our proposed algorithms can effectively detect the gradual shot changes based on the Class 1 information. Compared to our class information, the method based on the *Intra MB number*, *SMC*(*t*) and *SMC*$_{Intra\_included}$(*t*) have low effectiveness in detecting the gradual shot changes. We can see from Fig. 8 (b)-(d) that *the Intra MB number*, *SMC*(*t*) and *SMC*$_{Intra\_included}$(*t*) have similar values for frames inside and outside the shot change period. This makes them very difficult to differentiate the gradual-shot-change frames. Fig. 8 (c) shows that *SMC*(*t*) is the least effective. This implies that only using motion smoothness information cannot work well in detecting shot changes. Our experiments show that the effectiveness of *SMC*(*t*) will be further reduced when both of the sub-sequences before and after the shot change have similar patterns or low motions. In these cases, the motion unsmoothness will not be so obvious at the shot change.

Table 1 compares the *Miss* rate, the *False Alarm* rate, and the total error frame rate (TEFR) [6] for different algorithms in detecting the shot changes in an extended TRECVID dataset. The extended TRECVID dataset has totally 60 sequences which include both the sequences from the public TRECVID dataset [28-29] and the sequences that we create. There are totally 16 abrupt shot change sequences and 62 gradual shot change sequences with different types (gradual transfer, fade-in and fade-out) and with different length of shot-changing period (e.g., 10 frames, 20 frames, and 30 frames). The example sequences of the dataset are shown in Fig. 9. The *Miss* rate is defined by $N^k_{miss} / N^k_+$, where $N^k_{miss}$ is the total number of mis-detected shot change frames in sequence *k* and $N^k_+$ is the total number of shot change frames in sequence *k*. The *False Alarm* rate is defined by $N^k_{FA} / N^k_-$, where $N^k_{FA}$ is the total number of *false alarmed* frames in sequence *k* and $N^k_-$ is the total number of *non-shot-change* frames in sequence *k*. We calculate the *Miss* rate and the *False Alarm* rate for each sequence and average the rates. The Total Error Frame Rate (*TEFR*) rate is defined by $N_{t\_miss} / N_{t\_f}$, where $N_{t\_miss}$ is the total number of mis-detected shot change frames for all sequences and $N_{t\_f}$ is the total number of frames in the dataset. The *TEFR* rate reflects the overall performance of the algorithms in detecting all sequences.

In order to have a fair comparison, we also list the results of only using Class 1 information for detection (i.e., detect a shot change frame if $N_{class\_1}(t) < T_1$, *Proposed-Class 1 only* in Table 1). In the experiments of Table 1, the thresholds for detecting shot changes in Method (1) (*Intra-based*), Method (2) (*MV-Smooth-based*) and Method (3) (*Intra+MV_Smooth*) are set to be *200*, *2000* and *105000*, respectively. These thresholds are selected based on the experimental statistics.

Table 1 Performance comparison of different algorithms in detecting the shot changes in the extended TRACVID dataset

|  | Miss (%) | False Alarm (%) | TEFR |
|---|---|---|---|
| *Intra-based* | 25.24 | 4.27 | 13.50 |
| *MV-Smooth-based* | 43.72 | 17.36 | 22.75 |
| *Intra+MV-Smooth* | 24.71 | 3.49 | 12.58 |
| *Proposed-Class 1 only* | 8.34 | 3.81 | 5.51 |
| *Proposed-All Class+Intra* | 6.13 | 2.91 | 3.23 |

From Table 1, we can see that the performances of our proposed algorithms (*Proposed-Class 1 only* and *Proposed-All Class+Intra*) are better than the other methods.

Furthermore, several other observations can be drawn from Table 1 as follows:

(1) Basically, our Class 1 information, the Intra MB information [26-27] and the residue information [30] can all be viewed as the features to measure the content correlation between frames. However, from Table 1, we can see that the performance of our *Proposed-Class 1 only* method is obviously better than the *Intra-based* method. This is because the Class 1 information includes both the residue information and the motion information. Only those MBs with both regular motion patterns (i.e., *MV* close to *PMV* or (0,0) *MV*) and low-matching-cost values are classified as Class 1. We believe that these MBs can reflect more efficiently the nature of the content correlation between frames. In our experiment, we found that there are a large portion of MBs in the gradual-shot-change frames where neither intra nor inter prediction can perform well. The inter/intra mode selections for these MBs are quite random, which affects the performance of the *Intra-based* method. Compared to the *Intra-based* method, our algorithm can work well by simply classifying these MBs outside Class 1 and discarding them from the shot change detection process.

(2) The performance of the *Proposed-All Class+Intra* method can further improve the performance from the *Proposed-Class 1 only* method. This implies that including Class 2 and Class 3 can help detect those frames that cannot be easily differentiated by only using the Class 1 information at the boundary of the shot change period. Furthermore, the reduced FA rate of the *Proposed-All Class+Intra* method also implies that including the intra-coded MB information can help discard false alarm frames due to MB misclassification.

For further demonstrating the effectiveness of our class information, we conduct another experiment by utilizing the linear support vector machine (linear SVM) [22] for shot change detection (i.e., extracting features for each frame and then using SVM to detect shot changes). As mentioned, the advantage for using SVM is that the decision rules can be automatically obtained from the training process instead of using the manually set rules in Eqn. (3) [21]. In this experiment, we compare our class information with three recently proposed features for shot change detection. They are as follows:

(1) Inter prediction mode information [21]. (*Inter-mode+SVM* in Table 2)

(2) Local indicators [16] (*Local-Indi+SVM* in Table 2)

(3) Color feature and reliable MV proportions [31] (*Color+relyMV+SVM* in Table 2)

(4) Our proposed class information and Intra information (*Proposed-All Class+Intra+SVM* in Table 2)

Note that in order to have a fair comparison, only the features are borrowed from the reference works [16, 21, 31] while all the decision rules are obtained by training the SVM. The shot change detection results and the average operation time (AOT, the average operation time for performing shot change detection on each frame) is shown in Table 2.

Table 2 Performance comparison by using different features in the extended TRACVID dataset when SVN is used for shot change detection

|  | Miss (%) | False Alarm (%) | TEFR | AOT(ms) |
|---|---|---|---|---|
| *Inter-mode+SVM* | 12.93 | 6.04 | 8.23 | 3.5 |
| *Local-Indi+SVM* | 7.68 | 3.27 | 4.57 | 5.6 |
| *Color+relyMV+SVM* | 4.28 | 2.30 | 2.91 | 37.8 |
| *Proposed-All Class+Intra+SVM* | 4.72 | 2.15 | 2.97 | 2.3 |

Several observations can be obtained from Table 2: **(a)** Comparing Table 2 with Table 3, we can see that the performance of our class information (*Proposed-All Class+Intra+SVM*) is improved. It demonstrates that SVM can achieve more sophisticated decision rules than our manually set rules in Eqn. (3). **(b)** Using inter prediction mode information only (i.e., *Inter-mode+HMM*) have less satisfactory results since they do not include MV information. Similarly, although local indicator features (*Local-Indi+HMM*) can effectively detect abrupt changes, they are less effective in gradual shot changes due to the lack of MV information. Compared to these two methods, the reliable MV proportion method (*Color+relyMV+HMM*) as well as our proposed class information (*Class+Intra+HMM*) can achieve better shot detection results. (c) Although the reliable MV proportion method has the best performance, its complexity is high. Compared to this, our proposed class information can achieve similar performance while with obviously low complexities.

## B. Experimental Results for Motion Discontinuity Detection

In this section, we perform experiments for MD detection. The following four methods are compared. Method (1)-(3) are the same as the previous section.

**(1)** Detect MD based on the number of Intra MBs (*Intra-based*).

**(2)** Detect MD based on motion smoothness (*MV-Smooth-based*).

**(3)** Detect MD based on the combined information of Intra MB and motion smoothness (*Intra+MV-Smooth*).

**(4)** Our proposed MD detection algorithm as in Eqn. (5) (*Proposed*).

Fig. 10 shows the curves of features that are used in the above algorithms for *Stefan_Sif* sequence. Fig. 10 (a) shows the ground truth segment of Smooth Camera Motions. In Fig. 10 (a), the segments valued *0* represent SCMs with low or no camera motion and the segments with value *1* represent SCMs with high or active camera motion. For example, the segment between frame *177* and *199* represents an SCM where there is a rapid rightward of the camera; and the segment between frame *286* and *300* represents an SCM of a quick zoom-in of the camera. The frames between SCMs are the Motion Discontinuity frames that we want to detect. The ground truth MD frames are labeled as the vertical dashed lines in Fig. 10 (b)-(e). It should be noted that most MDs in Fig. 10 include several frames instead of only one. Fig. 10 (b)-(e) show the curves of *the number of Intra MBs*, $SMC(t)$, $SMC_{Intra\_included}(t)$, and *the number of Class 2 MBs*, respectively.

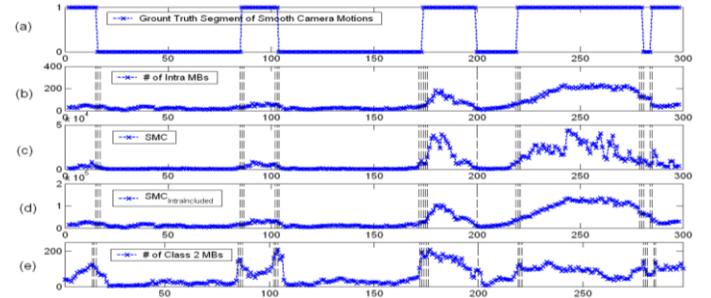

Fig. 10 Feature curves for the MD detection in *Stefan_Sif*.

Several observations can be drawn from Fig. 10 (b)-(e) as follows:

(1) Our Class 2 information is more effective in detecting the MDs. For example, in Fig. 10-(e), we can see that our Class 2 information has strong response when the first three MDs happen. Comparatively, the other features in Fig. 10 (b)-(d) have low or no response. This implies that Method (I)- (III) will easily miss these MDs.

(2) Our Class 2 information has quicker and sharper response to MDs. For example, the value of our Class 2 information increases quickly at the places of the fourth (around frame *175*) and sixth (around frame *220*) MDs, while the other features response much slower or more gradual.

(3) Fig. 10 also demonstrates that our Class 2 information is a better measure of the motion unsmoothness. Actually the largest camera motion in *Stefan_Sif* takes place in the segment between frame *222* and frame *278*. However, we can see from Fig. 10-(e) that the values of the Class 2 information are not the largest in this period. This is because although the camera motion is large, the motion pattern is pretty smooth during the period. Therefore, a big number of MBs will have regular and predictable motions and will not be classified as Class 2. In most cases, our Class 2 information will have the largest responses when the motion pattern changes or the motion smoothness disrupts. Compared to our Class 2 information, other features are more sensitive to the 'motion strength' rather than the 'motion unsmoothness'. Furthermore, although SMC can also be viewed as a measure of the motion smoothness, we can see from Fig. 10 that our Class 2 information is obviously a better measure for motion unsmoothness.

Fig. 11-(b) shows the MD detection result of the *proposed* method based on the Class 2 information in Fig. 10-(e), where $k$, $th_1^{shot}$ and $th_3^{shot}$ in Eqn. (5) are set to be *4*, *50* and *100*, respectively. From Fig. 11, we can see that: **(a)** the *proposed* method can detect most MDs except the one at frame *200*. The *frame-200* MD is missed because we use a large window size of 5 frames (i.e., $k=4$ in Eqn. (5)). This MD can also be detected if we select a smaller window size. **(b)** Since the *proposed* method detects MDs based on the information of several frames, some delay may be introduced. We can see that the MDs detected in Fig. 11-(b) have a delay of a couple of frames from the ground truth in Fig. 11-(a). **(c)** There are also some false alarms such as the period between frame *180* and *190*. This is because the camera motions in these periods are too rapid. In these cases, the motion prediction accuracy will be decreased and some irregular global motions will be included. These factors will prevent the number of Class 2 MBs from decreasing after the *MD* finish. In these cases, some post-processing steps may be needed to discard these false alarms.

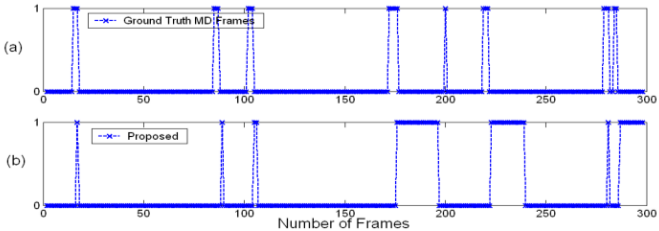

Fig. 11 The detection result of the *proposed* algorithm in *Stefan_Sif*.

As another example, Fig. 12 shows the feature curves for the *Coastguard_Cif* sequence, respectively. In this sequence, there are four obvious MDs: the first two belongs to a rapid upward of the camera and the second two belong to the small shakes of the camera. From Fig. 12, we can further see the effectiveness of our MB class information: our Class 2 information can effectively detect the two camera-shake MDs while the other methods will easily miss them. This is because when the magnitude of camera shake is small, the MV difference between frames will also be small, thus resulting in a small SMC. Furthermore, since the motion compensation still perform well in case of small camera shakes, the number of Intra MBs will also change little. However, our Class 2 information will effectively respond to these small camera shakes by classifying a large number of motion-unpredictable MBs into Class 2.

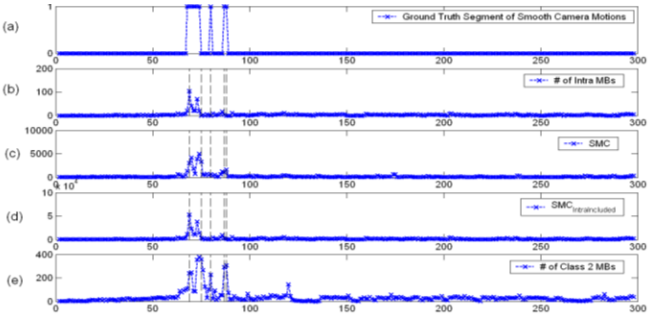

Fig. 12 Feature curves for the *MD* detection in *Coastguard_Cif*.

## C. Experimental Results for Global Motion Estimation

We compare the following four GME algorithms. For all of the methods, we use the same 6-parameter model for estimating the global motions, as in Eqn. (7)

**(1)** Do not discard the foreground MBs and directly use the Lease-Square method [17] to estimate the global model parameters (LS-6)

**(2)** Use the MPEG-4 VM global motion estimation method [32] (MPEG-4)

**(3)** Use the method in [17] for global motion estimation. In [17], an MV histogram is constructed for parameter estimation to speed up the global motion estimation process (MSU)

**(4)** Use the method in [12] to segment and discard foreground MBs and perform GME on the background MBs (P-Seg)

**(5)** Use our MB class information to segment and discard foreground MBs and perform GME on the background MBs, as described in Section III-C (Proposed)

Table 3 compares the Mean Square Error (MSE) of the global motion compensated results of the five algorithms. Normally, a small MSE value can be expected if the global motion parameter is precisely estimated. Table 4 compares the average MSE and the average operation time for different methods. Furthermore, Fig. 13 also show the subjective global-motion-compensated results for the five methods.

Table 3 Comparison of global-motion-compensated MSE results for different GME methods.

|            | LS-6  | MPEG-4 | MSU   | P-Seg | Proposed |
|------------|-------|--------|-------|-------|----------|
| Bus        | 27.73 | 22.67  | 22.85 | 23.42 | 22.32    |
| Stefan     | 22.71 | 20.99  | 19.09 | 19.36 | 19.52    |
| Flowertree | 24.92 | 20.66  | 21.51 | 20.83 | 19.72    |

Table 4 Comparison of average MSE and average operation time for different GME methods (Note: the operation time for the object segmentation part for P-Seg is taken from [12])

|                             | LS-6  | MPEG-4 | MSU   | P-Seg | Proposed |
|-----------------------------|-------|--------|-------|-------|----------|
| Average MSE                 | 25.12 | 21.44  | 21.15 | 21.20 | 20.52    |
| Average operation time (ms) | 17    | 376    | 56    | 37    | 25       |

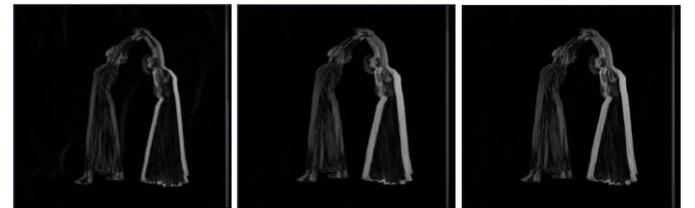

(a) LS-6      (b) MPEG-4      (c) MSU

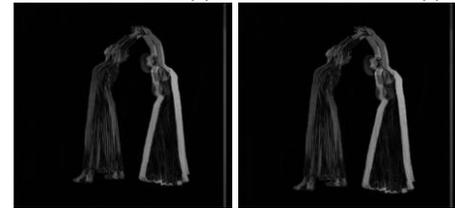

(d) P-Seg      (e) Proposed

Fig. 13 Subjective global-motion- compensated results of the four methods for *Dancer_cif*.

Some observations can be drawn from Table 3-4 and Fig. 13 as follows:

**(a)** Since the LS-6 method does not differentiate foreground and background, it cannot estimate the global motion of the background precisely. We can see from Table 3 that the LS-6 method has larger MSE values. Furthermore, Fig. 13-(a) also shows that there are obvious background textures in the compensated frame.

**(b)** Compared to the LS-6 method, the other four methods will segment and discard the foreground MBs before estimating the global motion for the background. We can see that our proposed method can achieve similar performance to the MEPG-4 and MSU methods.

**(c)** Since the MEPG-4 algorithm uses a three-layer method to find the outlier (foreground) pixels, its computation complexity is high. Although the MSU and the P-Seg algorithms reduce the complexity by constructing histograms or performing volume growth for estimating the foreground area, they still requires several steps of extra computations for estimating the global parameters. Compared with these two methods, our proposed method segments the foreground based on the readily available class information, the extra computation complexity is obviously minimum. Note that this operation time reduction will become very obvious and important when the GME algorithms are integrated with the computation-intensive video compression module for real-time applications.

**(d)** Although P-Seg can create good object segmentation results, its GME performance is not as good as our method. This is because our proposed algorithm focuses on detecting and filtering the "irregular motion" blocks while P-Seg more focuses on segmenting a complete object. By using our algorithm, blocks which do not belong to the foreground but have irregular motions will also be filtered from the GME process. This further improves the GME performance.

V. DISCUSSION AND ALGORITHM EXTENSION

In this section, we discuss some additional advantages and possible extensions of the algorithm. They are described in the following.

(1) It should be noted that we only discuss some example applications of our MB class information in this paper. We believe that our proposed class information can be used in many other video processing applications. For example, the MB class information can be used for rate control where the total number of MBs in each class can be used for frame-level bit allocation and the class label of each MB can be used for MB-level bit allocation. Similarly, we can also use the proposed MB class information for computation control motion estimation or rate control [8].

(2) As mentioned, the idea of our MB class information is general and it can be easily extended in different ways. For example, we can define more classes instead of three to have a more precise description of the frame content. We can also extend our MB class information to multiple partition sizes or multiple reference frame cases [1].

VI. SUMMARY

In this paper, a new MB class information is proposed for various video processing applications. We first propose to classify Macroblocks of each frame into different classes and use this class information to describe the frame content. Based on the proposed method, we further propose several algorithms for various video processing applications including shot change detection, motion discontinuity detection and global motion estimation. Experimental results demonstrate that methods based on the proposed class information can work efficiently and perform better than many of the existing methods.